\documentclass{PoS}
\usepackage{scalefnt}
\usepackage{epsf,amsmath,amssymb,graphicx,dcolumn}

\newcommand{\abbrev}{\rm\scalefont{.9}}
\newcommand{\lhc}{{\abbrev LHC}}
\newcommand{\pt}{\ensuremath{p_T}}
\newcommand{\qt}{\ensuremath{p_T}}

\newcommand{\nll}{{\abbrev NLL}}
\newcommand{\nnll}{{\abbrev NNLL}}
\newcommand{\lo}{{\abbrev LO}}
\newcommand{\nlo}{{\abbrev NLO}}
\newcommand{\nnlo}{{\abbrev NNLO}}
\newcommand{\sm}{{\abbrev SM}}
\newcommand{\qcd}{{\abbrev QCD}}
\newcommand{\thdm}{{\abbrev 2HDM}}
\newcommand{\mssm}{{\abbrev MSSM}}
\newcommand{\fs}[1]{#1{\abbrev FS}}
\newcommand{\mb}{m_{\rm b}}
\newcommand{\mhiggs}{M}
\newcommand{\mgen}{M}
\newcommand{\bbh}{b\bar bH}

\newcommand{\plus}{{\abbrev +}}
\newcommand{\citere}[1]{Ref.\,\cite{#1}}
\newcommand{\citeres}[1]{Refs.\,\cite{#1}}
\newcommand{\eqn}[1]{Eq.\,(\ref{#1})}

\newcommand{\fig}[1]{Fig.\,\ref{#1}}

\newcommand{\sct}[1]{Section~\ref{#1}}

\newcommand{\dd}{{\rm d}}
\newcommand{\dy}{{\abbrev DY}}
\newcommand{\ccbar}{c\bar c}

\newcommand{\als}{\ensuremath{\alpha_s}}
\newcommand{\bob}{\ensuremath{b_0/b}}
\newcommand{\bobsq}{\ensuremath{b^2_0/b^2}}

\newcommand{\hardcoef}[2]{H_{#1}^{#2}}

\newcommand{\hardh}[1]{\hardcoef{b}{H#1}}

\newcommand{\api}{\frac{\alpha_s}{\pi}}
\newcommand{\muF}{\mu_{\rm F}}
\newcommand{\muR}{\mu_{\rm R}}
\newcommand{\Qres}{Q}

\newcommand{\msbar}{\overline{\text{MS}}}

\newcommand{\pdf}{{\abbrev PDF}}

\newcounter{notecount}
\title{Resummed Higgs $p_T$ distribution at 
\nnlo{}\plus\nnll{}
in bottom-quark annihilation}

\ShortTitle{Resummed Higgs $p_T$ distribution at 
\nnlo{}\plus\nnll{}
in bottom-quark annihilation}

\author{Robert Harlander\\
        Fachbereich C, Bergische Universit\"at Wuppertal, Germany\\
        E-mail: \email{robert.harlander@uni-wuppertal.de}}

\author{\speaker{Anurag Tripathi} \\
  Universit\`{a} di Torino and INFN, Sezione di Torino, Italy \\ E-mail:
  \email{tripathi.hri@gmail.com}}

\author{Marius Wiesemann\\
        Physik-Institut, Universit\"at Z\"urich, Switzerland\\
        E-mail: \email{dr.wiesemann@gmail.com}}
\abstract{
The resummed transverse-momentum distribution for Higgs bosons produced
via bottom-quark annihilation at the \lhc{} is presented. Our results
are obtained in the five-flavor scheme to \nnlo{}\plus\nnll{}
accuracy. We present a theoretical prediction which consistently matches
the cross section at small and large transverse momenta. Theoretical
uncertainties are derived from a variation of the unphysical scales
entering the calculation. Their size is significantly reduced with
respect to lower orders.
}

\FullConference{Loops and Legs in Quantum Field Theory\\
                 27 April 2014 - 02 May 2014\\
                 Weimar, Germany}

\begin{document}

\section{Introduction}

After the discovery of a Higgs boson
\cite{Aad:2012tfa,Chatrchyan:2012ufa}, differential quantities to
determine its precise properties have become crucial. While in the
Standard Model (\sm{}), Higgs boson production proceeds predominantly
through gluon fusion, in theories with extended Higgs sector, such as
the Two-Higgs-Doublet Model (\thdm{}) or the Minimal Supersymmetric
\sm{} (\mssm{}), the associated production of a Higgs with bottom quarks
($\bbh{}$) can become the dominant Higgs production mechanism due to an
enhancement of the bottom Yukawa coupling (see
Refs.\,\cite{Dittmaier:2011ti,Dittmaier:2012vm,Heinemeyer:2013tqa} and
references therein). Two complementary approaches have been pursued in
the literature to obtain a theoretical prediction for this process.  In
the so-called {\it four-flavor scheme} (\fs{4}), which is most suitable
when final-state bottom quarks are considered as part of the signature,
the partonic processes at leading order (\lo{}) are $q\bar q\to b\bar
bH$ and $gg\to b\bar bH$ ($q\in\{u,d,c,s\}$). The cross section is known
up to next-to-\lo{} (\nlo{}) \qcd{}
\cite{Dittmaier:2003ej,Dawson:2005vi,Dawson:2003kb}.  In the {\it
  five-flavor scheme} (\fs{5}) the \lo{} partonic process is $b\bar b\to
H$. The final state bottom quarks are implicitly integrated out in the
parton model, which, therefore, are considered as being remnants of the
proton. This approach is most suitable for the calculation of the $\bbh$
component to inclusive Higgs production, since it implicitely resums
collinear logarithms through {\abbrev DGLAP} evolution. The inclusive
total cross section in the \fs{5} is known at next-to-\nlo{}
(\nnlo{})~\cite{Harlander:2003ai}.

Kinematical distributions of the Higgs boson will become more and more
important to decrease the uncertainties on the measured Higgs properties
and the search for possible deviations from the \sm{} predictions. The
transverse momentum ($p_T$) spectrum of the Higgs is one of the first
differential quantities where the experimental analysis will receive
sensitivity, once sufficient luminosity is available.  In gluon fusion,
the $\pt$ distribution has been studied in great detail (see
Ref.\,\cite{Dittmaier:2012vm}). For $\bbh$, the spectrum of the Higgs
for $\pt>0$ is known at \nnlo{} in the \fs{5}
\cite{Harlander:2010cz,Ozeren:2010qp,Harlander:2011fx,%
  Buehler:2012cu}.\footnote{Throughout this paper we consistently refer
  to $b\bar{b}\rightarrow H$ as the \lo{} process, even though its
  contribution is $\sim\delta(\pt)$.}
%The special case of $H+b$ production had been
%considered earlier in \citere{Campbell:2002zm}.
  Until recently, transverse momentum resummation of the Higgs produced
  in bottom-quark annihilation has been considered only at
  \nlo{}\plus\nll{}\,\cite{Belyaev:2005bs}. We report in this article on
  the first calculation of the resummed \nnlo{}\plus\nnll{} transverse
  momentum distribution in the \fs{5}\,\cite{Harlander:2014hya}.

The remainder of the paper is organized as follows: In
\sct{sec:elements}, we briefly outline the $\pt$ resummation
formalism for the production of a colorless final state and discuss
the required resummation coefficients for $\bbh$ production. 
This discussion includes our result for the second order
 {\it hard coefficient} which was the last missing resummation coefficient at
\nnlo{}\plus\nnll{}. In \sct{sec:outcalc}, we sketch the calculation of 
matched-resummed cross section and describe the checks
that have been performed on the implementation. We present  
$\pt$-distributions for the \lhc{} at a center-of-mass
energy of 8\,TeV.

\section{Transverse momentum resummation}\label{sec:elements}
In the calculation of the transverse momentum distribution of a
colorless particle of mass $\mgen$, large logarithms $\pt/\mgen$ arise
at every order in $\als$ for $\pt\ll\mgen$. They emerge from an
incomplete cancellation of soft and collinear divergences. Therefore the
na\"ive perturbative expansion in $\als$ is no longer valid as $\pt\to
0$. However, their all-order resummation is feasible due to the
factorization of soft and collinear radiation from the hard process.
The resummation formula for the cross section in the low-$\pt$ region is
well known \cite{Collins:1984kg,Catani:2000vq}\footnote{Note that
  resummation occurs in the impact parameter ($b$) space.}
\begin{equation}
\begin{split}
\label{eq:res}
\frac{\dd \sigma^{F,\text{(res)}}}{\dd \qt^2} = \tau \int_0^{\infty}
\dd b\, \frac{b}{2} \,J_{0}(b \qt) \, W^F(b,\mgen,\tau)\,.
\end{split}
\end{equation}
The Mellin transform\footnote{$ W^F_N(b,\mgen) = \int_0^1\dd\tau
  \tau^{N-1} W^F(b,\mgen,\tau)$} of $W^F$ with respect to the variable
$\tau=\mgen^2/S$
is given by
\begin{equation}
\begin{split}
 W^F_N(b,\mgen) &= \sum_{c}\hat\sigma^{F,(0)}_{\ccbar}\,
 H_c^F(\als) 
%\\ & \times \, 
\exp \left \{- \int_{\bobsq}^{\mgen^2} \frac{\dd
   k^2}{k^2} \Big[ A_c(\als(k)) \ln \frac{\mgen^2}{k^2} + B_c(\als(k)) \Big]
 \right \}\\ & \times \sum_{i,j} C_{ci,N}(\als(\bob)) \, C_{\bar
   {c}j,N}(\als(\bob) ) \, f_{i,N}(\bob) \, f_{j,N}(\bob)\,,
\label{eq:wn}
\end{split}
\end{equation}
with $b_0=2\exp(-\gamma_E)$, Euler's constant $\gamma_E= 0.5772\ldots$,
the Bessel function $J_0(x)$, the hadronic center-of-mass energy $S$,
and $\hat\sigma_{\ccbar}^{F,(0)}$ the Born-level cross section. The
superscript $F$ identifies process specific quantities; e.g. $F=\bbh{}$
for the $\bbh$ process and $F=$\,\dy{} for Drell-Yan. Unless indicated
otherwise, the renormalization and factorization scales have been set to
$\muF=\muR=\mgen$. The index $c$ runs over all parton flavors
$c\in\{u,d,s,c,b\,,g\}$ and their charge conjugates.  The functions
$f_{i,N}(q)$ denote the Mellin transform of the parton density functions
$f_{i}(x,q)$.

The {\it resummation coefficients} can be expanded in $\als$:
\begin{equation}
\begin{split}
%C_{ci,N}(\als) &= \delta_{ci}
C_{ci,N}
%(\als) 
= \delta_{ci}
+ \sum_{n=1}^{\infty} \left(\api  \right)^n C_{ci,N}^{(n)}\,,
\hspace{0.4cm}
%\qquad
X
%(\als) 
= \sum_{n=1}^{\infty} \left(\api \right)^n
X^{(n)}\,, 
\hspace{0.4cm}
%\\
%H_c^F(\als) &= 1+\sum_{n=1}^{\infty} \left(\api \right)^n H_c^{F,(n)}\,,
H_c^F
%(\als) 
= 1+\sum_{n=1}^{\infty} \left(\api \right)^n H_c^{F,(n)}\,,
\label{eq:rescoef}
\end{split}
\end{equation}
where $X\in\{ A_c, B_c\}$. At {\it leading logarithmic} ({\abbrev LL})
accuracy only $A_c^{(1)}$ is relevant. At {\it next-to-{\abbrev\it LL}}
(\nll) also $A_c^{(2)}$, $B_c^{(1)}$, $C_{ci}^{(1)}$, and
$\hardcoef{c}{F,(1)}$ are included and $A_c^{(3)}$, $B_c^{(2)}$,
$C_{ci}^{(2)}$, and $\hardcoef{c}{F,(2)}$ enter at next-to-\nll{}
(\nnll{}) \cite{Kodaira:1981nh,Catani:1988vd, Davies:1984hs,
  Catani:2012qa}.
%[\red{references missing}]. 
For a fixed {\it resummation scheme}\,\cite{Bozzi:2005wk}, the
coefficients $A_c$, $B_c$, and $C_{ci}$ are process independent, while
the entire process dependence is embodied in the {\it hard coefficient}
$H_c^F$ and the Born factor $\hat\sigma_{\ccbar}^{F,(0)}$.  We will work
in the \dy{} resummation scheme, where $\hardcoef{q}{\dy}\equiv 1$ to
all orders (the final results are independent of this choice, of
course).  

For quark-initiated processes, all resummation coefficients are known
through \nnll{} in this scheme; the process specific hard coefficient
$H_b^{\bbh}\equiv \hardh{}$ was evaluated recently up to second order
for the $\bbh$ process\,\cite{Harlander:2014hya}. Its evaluation
requires the knowledge of the purely virtual amplitude for the process
$\bbh$ which was calculated through \nnlo{} in
\citeres{Harlander:2003ai,Ravindran:2006cg}. Following
Ref.\,\cite{Catani:2013tia}, we derive $\hardh{,(1)} = 3\,C_F$ and
\begin{equation}
\begin{split}
\hardh{,(2)} &= C_F\bigg[
\left( \frac{321}{64} - \frac{13}{48} \pi^2 \right) C_F +
\left(-\frac{365}{288} + \frac{\pi^2}{12} \right) N_f
%\\&\qquad
+ \left(\frac{5269}{576} - \frac{5}{12} \pi^2 - \frac{9}{4}
\zeta_3 \right) C_A\bigg]\,.
\label{eq:H2b}
\end{split}
\end{equation}
This yields a numerical value of $\hardh{,(2)} = 10.52\ldots$. 
We also calculated $\hardh{,(2)}$ using and independent numerical approach. The two results
are in perfect agreement with each other \cite{Harlander:2014hya}. This serves as an important check of our calculation.

%- }}}

\section{Outline of the calculation and results}\label{sec:outcalc}

In order to combine the resummed cross section in the low-\pt{} region
and the fixed order cross section at high $\pt$ we follow the additive
matching procedure of \citere{Bozzi:2005wk} and define the
resummed-matched result as follows:
\begin{align}
\label{eq:match}
\left[\frac{\dd\sigma^F}{\dd\pt^2}\right]_{\text{f.o.}+\text{l.a.}}=
\left[\frac{\dd\sigma^F}{\dd\pt^2}\right]_{\text{f.o.}}
-\left[\frac{\dd\sigma^{F,\text{(res)}}}{\dd\pt^2}\right]_{\text{f.o.}}
+\left[\frac{\dd\sigma^{F,\text{(res)}}}{\dd\pt^2}\right]_{\text{l.a.}}\,.
\end{align}
The fixed order $\pt$ distribution $[\dd\sigma]_\text{f.o.}$ is known
analytically up to \nnlo{} \cite{Ozeren:2010qp}, which has been checked
numerically against the partonic Monte Carlo program for $H+$jet
production of \citeres{Harlander:2010cz,Harlander:2011fx}.\footnote{For
  more details see also \citere{Wiesemann:2012ij}.}
%which in turn has been validated by various related calculations \cite{Campbell:2002zm,Frixione:2002ik,Frederix:2009yq,Frixione:2010ra,Hirschi:2011pa}.
The logarithmic terms $[\dd\sigma^\text{(res)}]_\text{f.o.}$ are
obtained from the perturbative expansion of \eqn{eq:res} up to
\nnlo{}.\footnote{See in Eqs.~(72) and (73) of \citere{Bozzi:2005wk}.}
We verified numerically that $[\dd\sigma^\text{(res)}]_\text{f.o.}$ and
$[\dd\sigma]_\text{f.o.}$ are identical in the limit $\pt\rightarrow 0$.
The calculation of the resummed cross section
$[\dd\sigma^{\text{(res)}}]_{\text{l.a.}}$ at low $\pt$ was carried out
using a modified version of the program {\tt
  HqT}~\cite{Bozzi:2003jy,Bozzi:2005wk,deFlorian:2011xf}, which we
extended to quark-induced processes, and implement the resummation
coefficients of the $\bbh$ process.

In addition, the pursued \pt{} resummation formalism \cite{Bozzi:2005wk}
implies the following {\it unitarity constraint}:
\begin{align}
\int \dd\pt^2
\left[\frac{\dd\sigma^F}{\dd\pt^2}\right]_{\text{f.o.+l.a.}}
&\equiv\left[\sigma^F_\text{tot}\right]_{\text{f.o.}}.
\label{eq:unitarity}
\end{align}
This relation serves as an important cross-check on our implementation
of the matched-resummed distribution. We verified it at the per-mille
level for various values of the resummation, factorization, and
renormalization scales, separately at order $\alpha_s$ and $\alpha_s^2$,
and for the individual partonic sub-channels.

We present results for the \lhc{} at $8$ TeV center-of-mass
energy in the \sm{}. These results can also be used for neutral ({\abbrev CP} even and odd) Higgs production
within the \thdm{} through rescaling of the bottom Yukawa coupling, and, according to the studies of
\citeres{Dawson:2011pe,Dittmaier:2006cz}, also within the \mssm{}.
Our choice for the central factoriza\-tion and renormalization
scale is $\muF=\muR=\mu_0\equiv \mhiggs$; our default value for the
resummation scale is $\Qres=Q_0\equiv \mhiggs/4$. If not stated
otherwise, all numbers are obtained with the {\abbrev
MSTW2008}~\cite{Martin:2009iq} \pdf{} set.
The bottom mass is set to zero throughout the calculation,
except for the bottom-Higgs Yukawa coupling which we insert in the
$\msbar$-scheme at the scale $\muR$, derived from the input value
$\mb(\mb)=4.16$\,GeV.

\begin{figure}
\begin{center}
    \begin{tabular}{cc}
     \hspace{-0.45cm}
     \mbox{\includegraphics[height=.253\textheight]{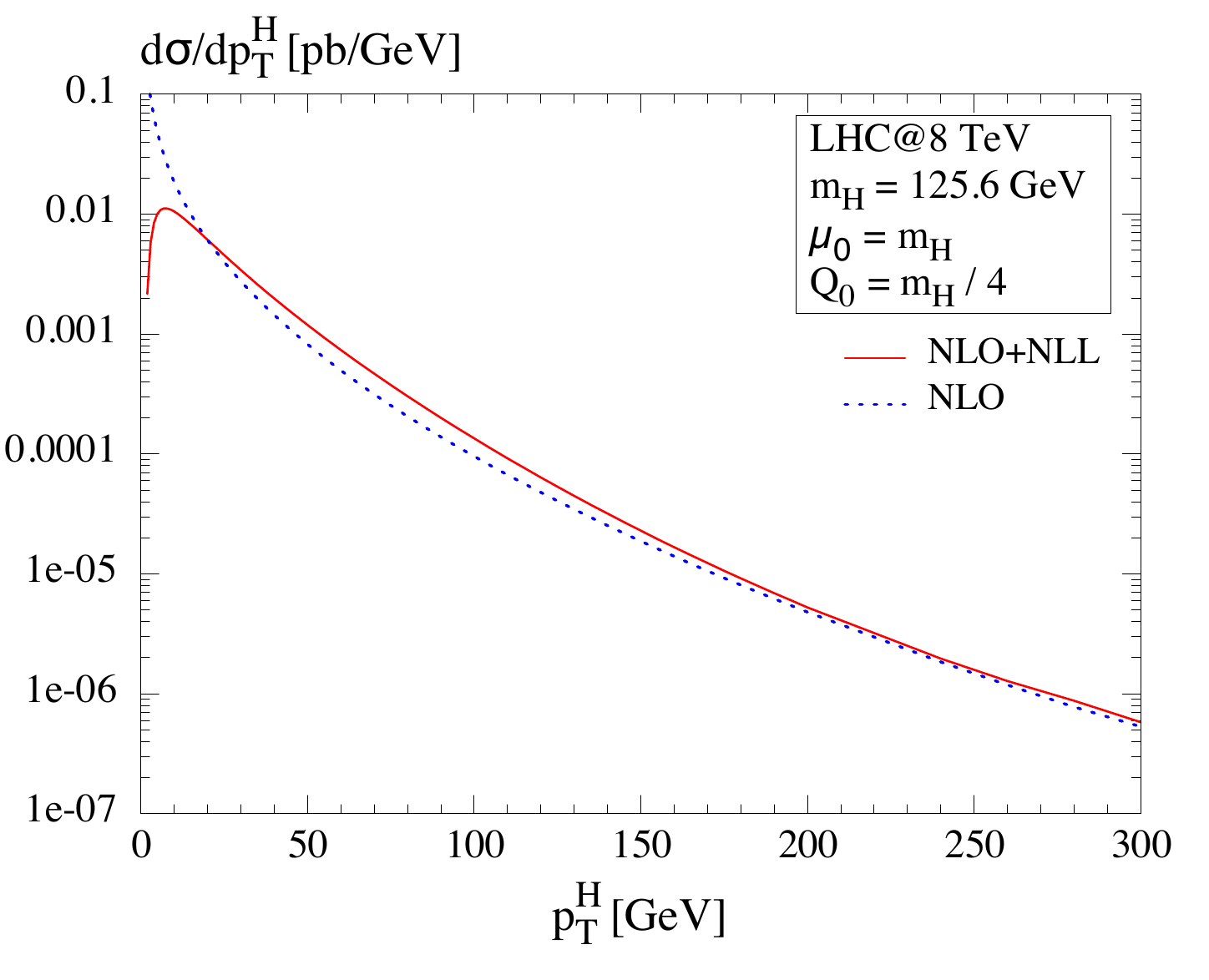}} &
     \mbox{\includegraphics[height=.253\textheight]{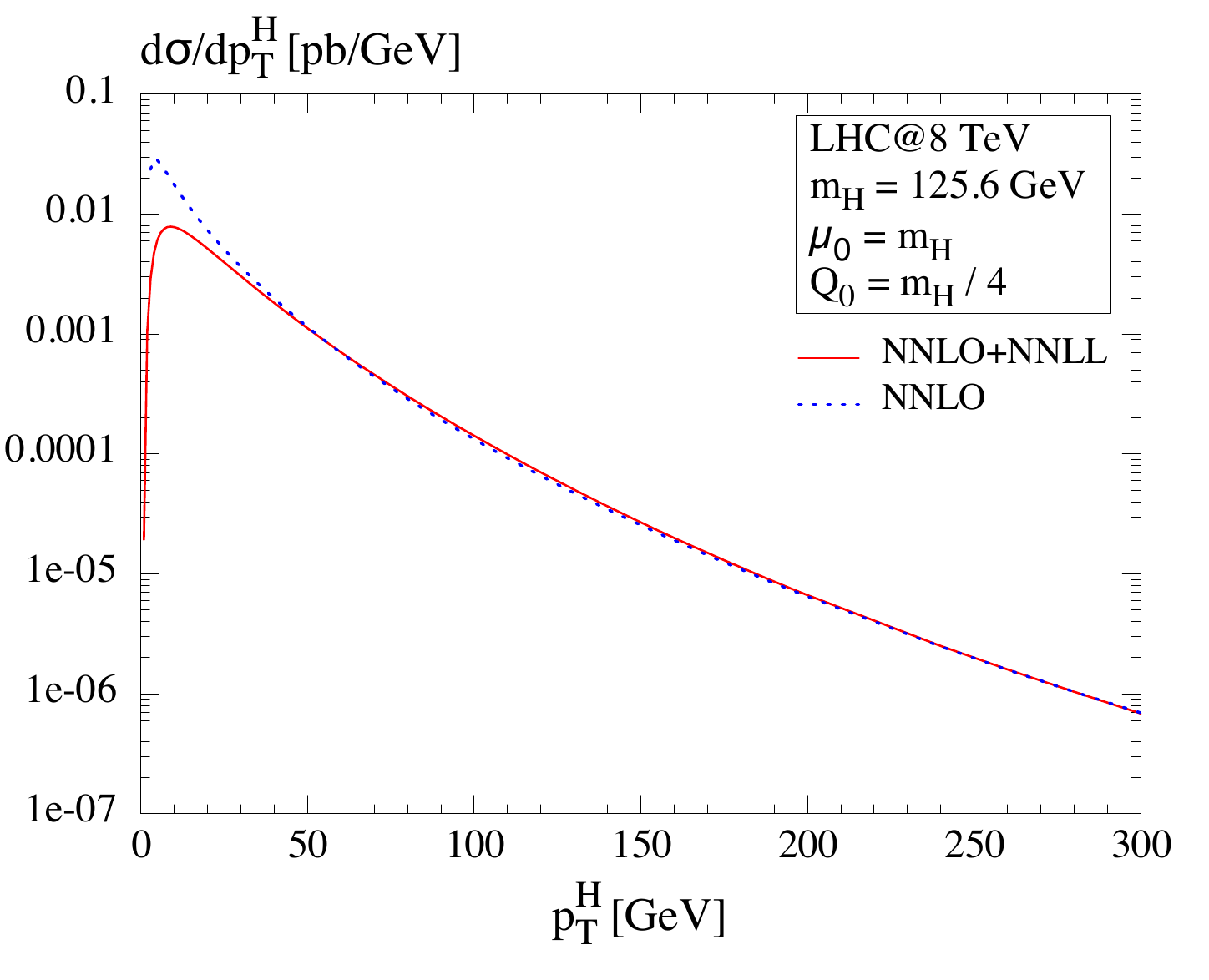}}
            \\
     \hspace{0.15cm} (a) & (b)
    \end{tabular}
    \parbox{.9\textwidth}{%
      \caption[]{\label{fig:highpT}{ Transverse momentum spectrum at
          (a) \nlo{} (blue, dashed) and \nlo\plus\nll{} (red, solid)
          and (b) \nnlo{} (blue, dashed) and \nnlo\plus\nnll{} (red, solid).
          (Here and in the following plots,
          $m_H=M$ is the Higgs mass.)}}}
\end{center}
\end{figure}
\fig{fig:highpT}\,(a) shows the \nlo{}\plus{}\nll{} together with the fixed
order \nlo{} distribution. Differences between the fixed-order and the resummed-matched
curve at $\pt\sim M$ are higher order effects \cite{Banfi:2013eda} which can be quite sizable
\,\cite{Belyaev:2005bs}, and were also observed for gluon fusion
\cite{Mantler:2012bj,Banfi:2013eda}. However, for our choice of the resummation scale $\Qres=M/4$ we observe a good high-\pt{} matching already at \nlo{}\plus{}\nll{}.
The agreement between fixed order
and resummed-matched curve is further improved at \nnlo{}\plus{}\nnll{}, 
see \fig{fig:highpT}\,(b). 
 The two curves are indistinguishable for $\pt\gtrsim 50$\,GeV.

%---------------
\begin{figure}
\begin{center}
    \begin{tabular}{cc}
     \hspace{-0.45cm}
     \mbox{\includegraphics[height=.253\textheight]{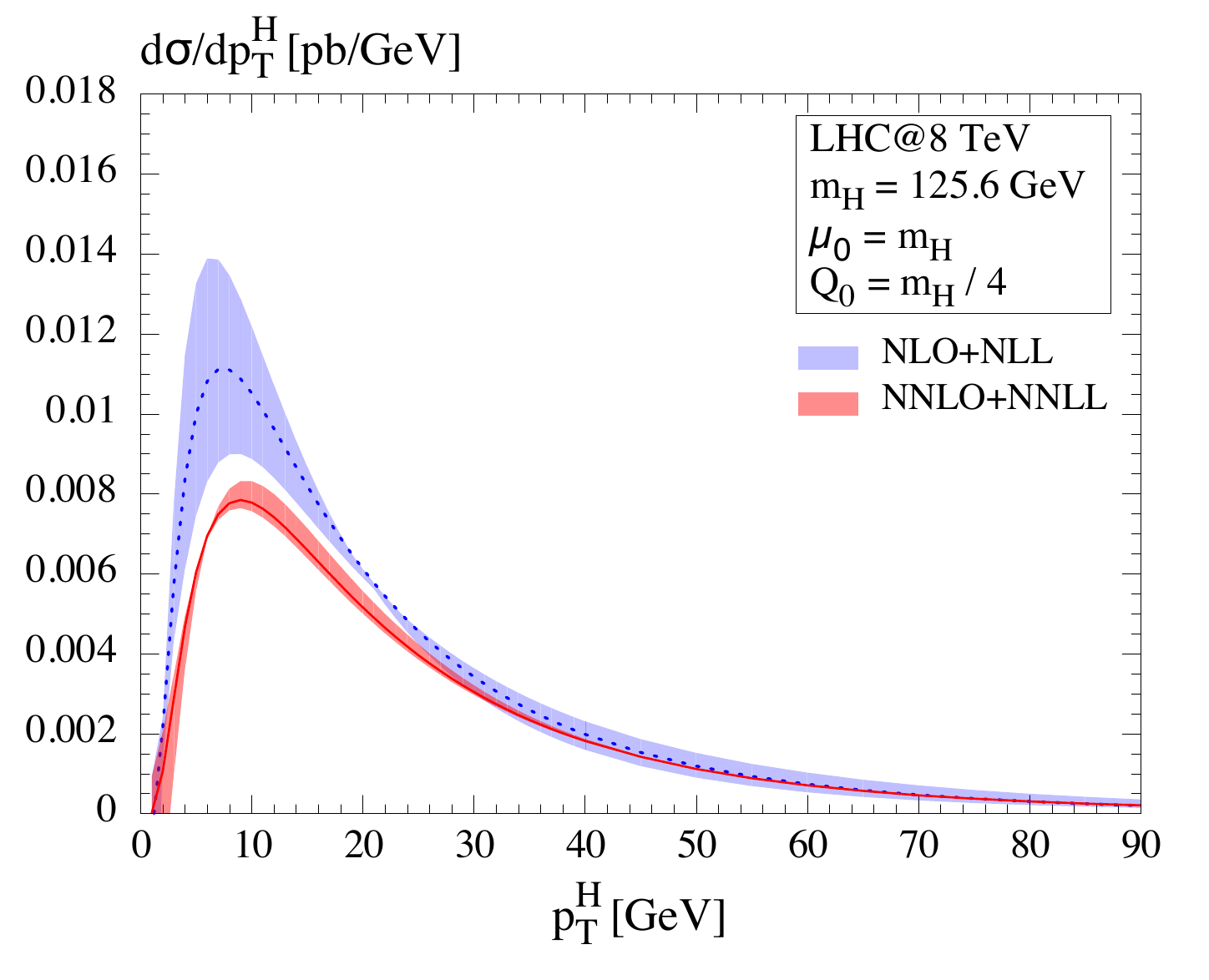} } &
     \mbox{\includegraphics[height=.253\textheight]{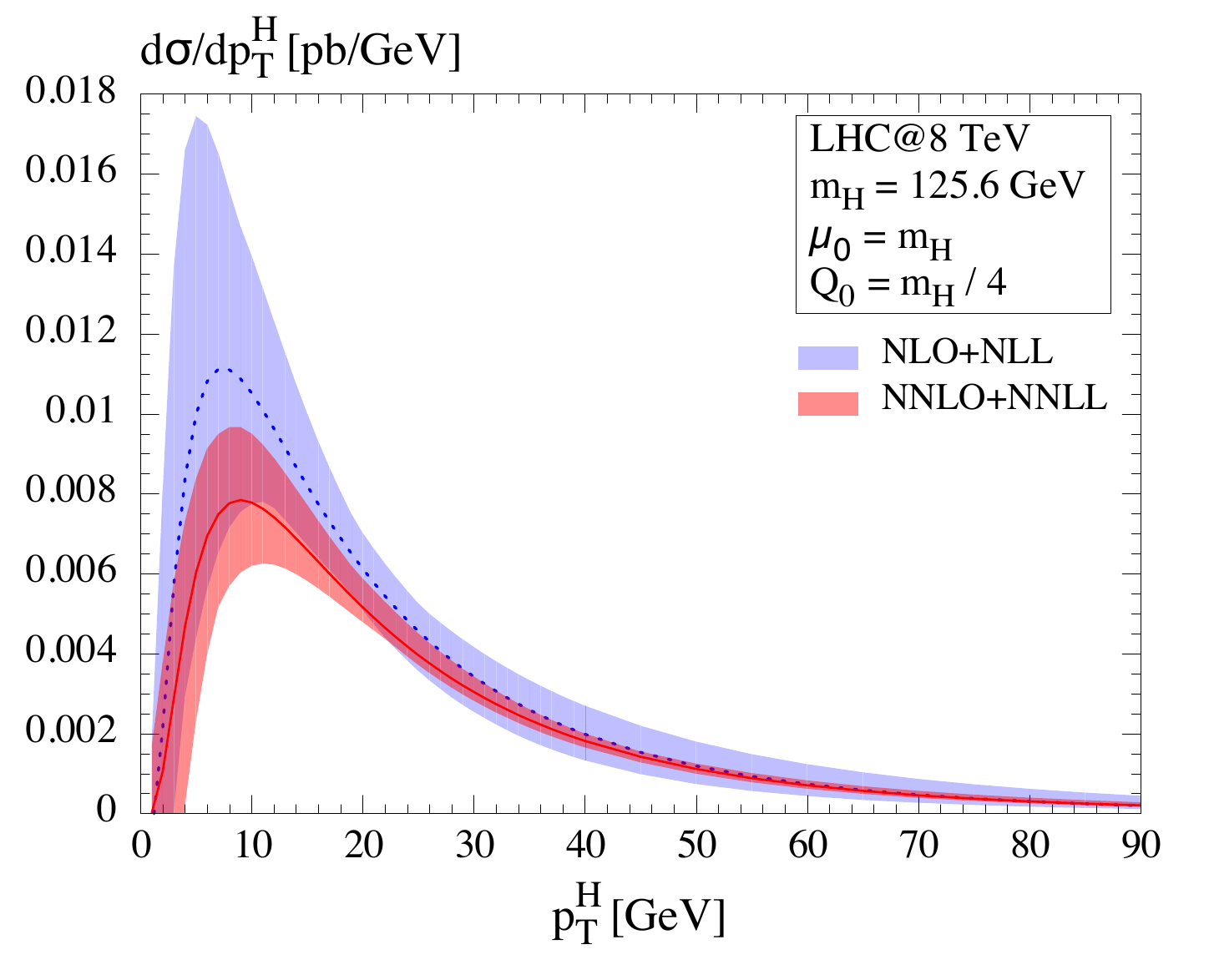}} 
            \\
     \hspace{0.15cm} (a) & (b)
    \end{tabular}
    \parbox{.9\textwidth}{%
      \caption[]{\label{fig:Qall}
{ Resummed-matched $p_T$-distribution
          at \nlo{}\plus\nll{} (blue, dashed) and \nnlo{}\plus\nnll{}
          (red, solid); lines: central scale choices; bands: (a) uncertainty
          due to $\Qres{}$-variation.  (b) uncertainty
          due to variation all scales.} }}
\end{center}
\end{figure}

Let us now consider the effect of inclusion of the higher orders in the
theoretical uncertainty.  At \nnlo{}\plus{}\nnll{}, a reduced
sensitivity of the cross section to the resummation scale is expected
compared to \nlo{}\plus{}\nll{}. This expectation is confirmed in
\fig{fig:Qall}\,(a). The bands indicate the uncertainties with respect
to $\Qres$, which are obtained by varying $\Qres$ between $Q_0/2$ and
$2\,Q_0$, while fixing $\muF$ and $\muR$ at their default values; the
lines correspond to $\Qres=Q_0$. In fact, the extremely mild
$\Qres$-dependence of the \nnlo{}\plus{}\nnll{} cross section is
remarkable.

Finally, \fig{fig:Qall}\,(b) shows the result for an independent variation of
all three unphysical scales within $\Qres\in[Q_0/2,2\,Q_0]$ and
$\muF,\muR\in[\mu_0/2,2\,\mu_0]$, while excluding the regions
$\muF/\muR > 2$ and $\muF/\muR < 1/2$. Again, we observe a
significant reduction of the scale uncertainties on the resummed-matched
 cross section at \nnlo{}\plus{}\nnll{} with respect to \nlo{}\plus\nll{}. 
At the maximum, the relative size of the uncertainty band is
 $+23/-23$\% for the \nnll{} curve and $+48/-41\%$
at \nll{}.
The corresponding plots for $13$ TeV and \pdf{} uncertainties are 
presented in \citere{Harlander:2014hya}.

\section{Conclusions}
We reported on the calculation of the missing second order hard
coefficient $\hardh{,(2)}$ and the resummed-matched \pt{} distribution
of Higgs bosons produced via bottom-quark annihilation at
\nnlo\plus\nnll. Our resummed low $\pt$ distribution matches well to the
fixed order distribution already around $50$\,GeV.  Furthermore, the
variation of the cross section with the unphysical scales is significant
reduced compared to \nlo{}\plus\nll{}. In fact, the extremely weak
dependence of the \nnlo{}\plus\nnll{} result on the resummation scale is
remarkable.  Our results provide a precise prediction that consistently
combines the cross section in the low- and high-$\pt{}$ region, which
should prove useful particularly in models with enhanced bottom Yukawa
coupling.

\paragraph{Acknowledgements.}
The work of AT was supported in part by the University of
Torino, the Compagnia di San Paolo, contract ORTO11TPXK, and DFG,
contract HA\,2990/5-1. MW was supported
by the European Commission through the FP7 Marie Curie Initial Training
Network ``LHCPhenoNet'' (PITN-GA-2010-264564) and BMBF, contract 05H12PXE.
%\appendix
%

\def\app#1#2#3{{\it Act.~Phys.~Pol.~}\jref{\bf B #1}{#2}{#3}}
\def\apa#1#2#3{{\it Act.~Phys.~Austr.~}\jref{\bf#1}{#2}{#3}}
\def\annphys#1#2#3{{\it Ann.~Phys.~}\jref{\bf #1}{#2}{#3}}
\def\cmp#1#2#3{{\it Comm.~Math.~Phys.~}\jref{\bf #1}{#2}{#3}}
\def\cpc#1#2#3{{\it Comp.~Phys.~Commun.~}\jref{\bf #1}{#2}{#3}}
\def\epjc#1#2#3{{\it Eur.\ Phys.\ J.\ }\jref{\bf C #1}{#2}{#3}}
\def\fortp#1#2#3{{\it Fortschr.~Phys.~}\jref{\bf#1}{#2}{#3}}
\def\ijmpc#1#2#3{{\it Int.~J.~Mod.~Phys.~}\jref{\bf C #1}{#2}{#3}}
\def\ijmpa#1#2#3{{\it Int.~J.~Mod.~Phys.~}\jref{\bf A #1}{#2}{#3}}
\def\jcp#1#2#3{{\it J.~Comp.~Phys.~}\jref{\bf #1}{#2}{#3}}
\def\jetp#1#2#3{{\it JETP~Lett.~}\jref{\bf #1}{#2}{#3}}
\def\jphysg#1#2#3{{\small\it J.~Phys.~G~}\jref{\bf #1}{#2}{#3}}
\def\jhep#1#2#3{{\small\it JHEP~}\jref{\bf #1}{#2}{#3}}
\def\mpla#1#2#3{{\it Mod.~Phys.~Lett.~}\jref{\bf A #1}{#2}{#3}}
\def\nima#1#2#3{{\it Nucl.~Inst.~Meth.~}\jref{\bf A #1}{#2}{#3}}
\def\npb#1#2#3{{\it Nucl.~Phys.~}\jref{\bf B #1}{#2}{#3}}
\def\nca#1#2#3{{\it Nuovo~Cim.~}\jref{\bf #1A}{#2}{#3}}
\def\plb#1#2#3{{\it Phys.~Lett.~}\jref{\bf B #1}{#2}{#3}}
\def\prc#1#2#3{{\it Phys.~Reports }\jref{\bf #1}{#2}{#3}}
\def\prd#1#2#3{{\it Phys.~Rev.~}\jref{\bf D #1}{#2}{#3}}
\def\pR#1#2#3{{\it Phys.~Rev.~}\jref{\bf #1}{#2}{#3}}
\def\prl#1#2#3{{\it Phys.~Rev.~Lett.~}\jref{\bf #1}{#2}{#3}}
\def\pr#1#2#3{{\it Phys.~Reports }\jref{\bf #1}{#2}{#3}}
\def\ptp#1#2#3{{\it Prog.~Theor.~Phys.~}\jref{\bf #1}{#2}{#3}}
\def\ppnp#1#2#3{{\it Prog.~Part.~Nucl.~Phys.~}\jref{\bf #1}{#2}{#3}}
\def\rmp#1#2#3{{\it Rev.~Mod.~Phys.~}\jref{\bf #1}{#2}{#3}}
\def\sovnp#1#2#3{{\it Sov.~J.~Nucl.~Phys.~}\jref{\bf #1}{#2}{#3}}
\def\sovus#1#2#3{{\it Sov.~Phys.~Usp.~}\jref{\bf #1}{#2}{#3}}
\def\tmf#1#2#3{{\it Teor.~Mat.~Fiz.~}\jref{\bf #1}{#2}{#3}}
\def\tmp#1#2#3{{\it Theor.~Math.~Phys.~}\jref{\bf #1}{#2}{#3}}
\def\yadfiz#1#2#3{{\it Yad.~Fiz.~}\jref{\bf #1}{#2}{#3}}
\def\zpc#1#2#3{{\it Z.~Phys.~}\jref{\bf C #1}{#2}{#3}}
\def\ibid#1#2#3{{ibid.~}\jref{\bf #1}{#2}{#3}}
\def\otherjournal#1#2#3#4{{\it #1}\jref{\bf #2}{#3}{#4}}
\newcommand{\jref}[3]{{\bf #1}, #3 (#2)}
\newcommand{\hepph}[1]{{\tt [hep-ph/#1]}}
\newcommand{\mathph}[1]{{\tt [math-ph/#1]}}
\newcommand{\arxiv}[2]{{\tt arXiv:#1}}
%\newcommand{\bibentry}[4]{#1, {\it #2}, #3\ifthenelse{\equal{#4}{}}{}{, }#4.}
%----------------------------------------------------------------------

\end{document}